# Apprentissage de la pensée informatique : de la formation des enseignant·e·s à la formation de tou·te·s les citoyen·ne·s.

Corinne Atlan[9], Jean-Pierre Archambault[10], Olivier Banus[9,5], Frédéric Bardeau[1], Amélie Blandeau[7], Antonin Cois[8], Martine Courbin[7], Gérard Giraudon[5,7], Saint-Clair Lefèvre[2], Valérie Letard[6], Bastien Masse[4], Florent Masseglia[7], Benjamin Ninassi[7], Sophie de Quatrebarbes[3], Margarida Romero[2,], Didier Roy[7,8], Thierry Viéville[2,7].

[1] Simplon.co https://simplon.cofred@simplon.co
[2] LINE http://unice.fr/laboratoires/linemargarida.romero@unice.fr
[3] S24B http://www.s24b.com sophiedequatrebarbes@gmail.com
[4] Université de Nantes bastien.masse@univ-nantes.fr
[5] Educazur https://educazur.fr
[6] La compagnie du code https://www.lacompagnieducode.org valerie.letard@lacompagnieducode.org
[7] Inria, Mission de Médiation Scientifique prenom.nom@inria.
[8] Poppy Station, https://www.poppystation.org
[9] Réseau Canopé, pôle national de compétence numérique éducatif https://www.reseau-canope.fr
[10] EPI https://www.epi.asso.fr

**Résumé.** En France au cours de ces dernières années, l'apprentissage de l'informatique (sous le terme d'« apprentissage du code ») est entré dans les programmes scolaires, en primaire et secondaire. Cet apprentissage vise notamment le développement de la pensée informatique (au sens défini par Wing) afin de permettre aux élèves, filles et garçons, d'acquérir les bases, une étape initiale vers la maîtrise du numérique, sous tous ses aspects (science, technologie, industrie et culture). Cependant, peu d'enseignant·e·s, ou de parents, ont été formé·e·s pour enseigner les sciences du numérique ou éduquer à leurs fondements et leurs usages. De plus, si le système éducatif avance progressivement au niveau de ces objectifs, dans la vie quotidienne et en contexte professionnel, il existe aussi un besoin de formation tout au long de la vie à la pensée informatique. Des projets d'envergure sur l'apprentissage du code sont aujourd'hui forts d'un véritable succès en matière de support à la formation des professionnel·le·s de l'éducation sur ces sujets. Cependant ces projets nécessitent une main d'oeuvre importante tant pour la création de ressources que pour leur actualisation, afin de rester en phase avec les besoins de formation dans un domaine en évolution constante. Dans le but de développer davantage les objectifs de démystification de la pensée informatique vers un large public de citoyens et de citoyennes, nous voulons questionner ici la manière dont il est possible de concevoir une initiative concrète et opérationnelle qui relève ce défi. Partageons ici une proposition et discutons-la. Ce qui est proposé porte un nom : une Université, Citoyenne en Sciences et Culture du Numérique (#UCscN) qui s'inscrit dans la tradition des universités populaires. Il s'agit donc très simplement d'étendre à toutes et tous cette éducation pour penser l'informatique en capitalisant sur l'expérience acquise de Class'Code en formant les professionnel·le·s de l'éducation.

**Mots-clés.** Pensée informatique, Égalité des chances, Parité, Fracture Numérique, Formation au long de la vie.

**Abstract.** In recent years, in France, computer learning (under the term of code) has entered the school curriculum, in primary and high school. This learning is also aimed at developing computer thinking to enable students, girls and boys, to start master all aspects of the digital world (science, technology, industry, culture). However, neither teachers, nor parents are trained to teach or educate on these topics. Furthermore, if the educational system progresses progressively towards these objectives, in everyday life and in professional context there is also a need for lifelong training in computer thinking. Large-scale projects on coding initiation are now quite successful in supporting the training of professionals in education on these topics. However, they require an infrastructure of people and important resources to maintain their level of efficiency. In order to further develop the objectives of helping people to demystify IT thinking, we aim to question here the way by which it is possible to conceive a concrete and operational initiative that addresses this issue. A huge challenge: Let's share a proposal here and discuss it.

# 1 Introduction

En France, alors que les instances dirigeantes ont récemment compris qu'il était urgent de ne plus attendre [1] pour initier nos filles et nos garçons à l'informatique et à la maîtrise du numérique, une majorité de personnes estiment ne pas avoir à se former au numérique [2], même si dans les faits elles constatent être bloqués au niveau des usages.
Le besoin de formation est immense. Sans une formation citoyenne à la littératie informatique, telle que définie en [4], impossible – par exemple – de développer le vote électronique sans provoquer une fracture de confiance vis à vis de qui ne maîtrise la technologie sous-jacente [5], d'assurer notre sécurité et notre souveraineté numérique [6], de décider collectivement des choix à faire en matière juridique ou même politique [7], d'accompagner durablement la transition numérique et la mobilité professionnelle [8], en particulier pour les mutations liées à ce que l'on appelle intelligence artificielle [3]. On pourra ainsi favoriser le développement d'une société plus équitable, où chaque citoyen peut participer autrement qu'en consommant des produits et services proposés par les géants de l'économie de plate-formes. En effet, une meilleure compréhension des enjeux et une meilleure diffusion des compétences à créer de nouveaux services numériques peut tempérer certains processus économiques comme l'ubérisation, [39], qui contourne les acteurs économiques « traditionnels » tout en polarisant le marché du travail vers accroissement de la proportion du travail non qualifié. Le numérique ne peut répondre aux besoins de la société, si seule une minorité détient les connaissances permettant de se forger une opinion et les compétences permettant d'agir. La création de ces services numériques plus équitables nécessite, entre autres, de savoir coder en ayant bien intégré l'impact du numérique sur la société. Cela est vrai y compris pour les décideur·e·s [42].

Une initiative comme Class'Code [9,10,11] a permis, avant la mise en place du CAPES d'informatique et sa généralisation [12], de commencer à former quelques dizaines de milliers de professionnel·le·s de l'éducation (animateur·rices·s, professeur·e·s).
Aujourd'hui, dans le prolongement de cette initiative, il s'agit de développer une véritable culture du numérique, c'est à dire d'apprendre à coder et décoder le numérique au sens de [40], prolongeant ainsi les actions de projets partenaires de Class'Code qui avaient déjà envisagé cet axe[11].

Nous avons fait face jusqu'à récemment à un manque d'études en sciences de l'éducation sur ces nouveaux apprentissages. Ce manque est aujourd'hui en passe d'être comblé [13], y compris au niveau des productions de Class'Code [14,15], par exemple avec la démarche d'initiation à la pensée informatique avec des activités débranchées [16,17]. Une revue récente de l'activité de recherche francophone sur l'apprentissage de la programmation montre l'intérêt croissant pour ces apprentissages [18]. De même, la didactique de l'informatique est une activité qui se développe en tant que discipline à la croisée des sciences de l'éducation et des sciences du numérique. Des ouvrages internationaux sur l'apprentissage de l'informatique (dont [19] disponible en français) permettent de disposer d'un cadre pour enseigner.

Il faudra toute une génération pour que les élèves éduqués progressivement à l'informatique deviennent des adultes formés sur ces sujets. Dans la vie quotidienne, pour les parents de ces élèves, les étudiant·e·s des disciplines non STIM[1], mais aussi les entreprises, il y a également un besoin immédiat de formation, un besoin de formation tout au long de la vie.
Est-il possible de s'appuyer sur le succès de Class'Code en matière de support à la formation des professionnel·le·s de l'éducation [20], pour concevoir une initiative concrète et opérationnelle qui relève ce défi de former citoyennes et citoyens ?

---
[1] STIM (ou STEM en anglais) : science, technologie, ingénierie et mathématiques, ce sont les disciplines qui ont dans leur curriculum une formation à l'informatique. https://fr.wikipedia.org/wiki/STEM_(disciplines)

## 2   Faire comprendre la nécessité de se former.

Une majorité de personnes estime ne pas avoir à se former au numérique, donc à l'informatique, et laisse l'initiative à leur employeur de le prévoir, pensant que les applications sont essentiellement professionnelles [2]. Cependant, aujourd'hui le numérique est présent dans tous les aspects de notre vie, et non plus seulement comme un outil de travail.

Pour exemple, les polémiques et les craintes actuelles au sujet du développement de l'intelligence artificielle montrent à quel point le besoin de formation est important [21]. Nous devons accompagner la prise de conscience d'un numérique certes pervasif, mais qui n'est ni magique, ni incontrôlable, et qui sera d'autant plus utile qu'il sera compris. Cette nécessité de formation est entendue par une minorité d'entreprises, comme Rakuten[25]. On peut citer le cas d'Orange [41] qui forme quel que soit le métier de ses employés, (également en matière de réinsertion [26]), y compris soutenues par l'État comme celles mentionnées dans [27].

Comment renverser cette tendance et convaincre chaque citoyen·ne de l'intérêt de se former à ces sujets, dans une société devenue largement numérique ?

La finalité est de permettre à chacun, quel que soit son âge, son niveau d'études, son statut social ou son genre de s'initier à la pensée informatique [38], et de développer les compétences[2] du XXI$^e$ siècle [28] telles que la pensée critique, créative, collaborative (voir Fig. 1) afin de maîtriser les compétences transversales essentielles à l'exercice de ses droits, à sa participation politique, économique, sociale et culturelle à l'ère du numérique [29] ; il est essentiel par exemple d'avoir les clés pour se positionner et débattre de sujets tels que l'émergence de ce qu'on nomme « intelligence artificielle » [3]. Pour faire face à un contexte numérique parfois opaque, l'exercice de la pensée critique et la capacité à s'engager créativement dans la résolution de problèmes permettent d'exercer une citoyenneté éclairée [36] par un usage du numérique éthique et respectueux [6]. L'apprentissage de la pensée informatique se veut émancipateur vis à vis des algorithmes qui régissent de plus en plus notre quotidien, et vise une plus grande exigence citoyenne envers les acteurs du numérique, politiques ou industriels qui mettent en place les environnements avec lesquels nous interagissons [11].

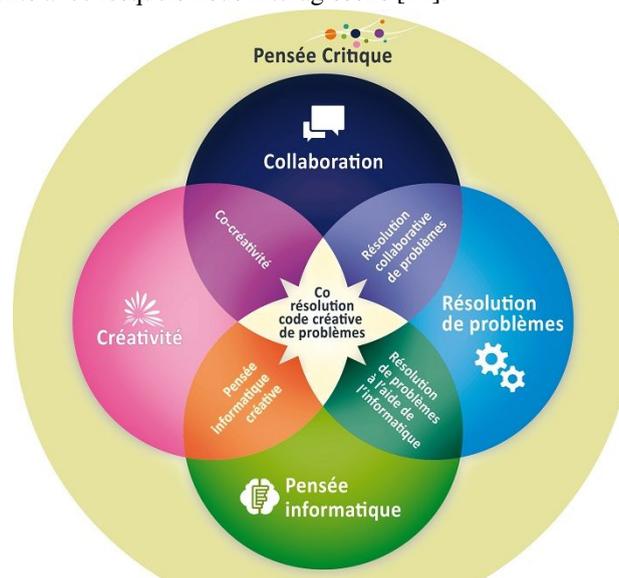

**Fig 1.** Les compétences du XXI$^e$ siècle en lien avec la pensée informatique [28].

---
[2]   Au sens de https://en.wikipedia.org/wiki/21st_century_skills

Voici les compétences que nous souhaitons partager à travers cette démarche :
- se construire une représentation opérationnelle du fonctionnement des algorithmes et de leur impact sur les différentes activités humaines où ils sont utilisés [3] ;
- disposer d'une culture minimale sur le sujet, pour s'y familiariser et pouvoir relier ces savoirs à notre propre culture [11] ;
- faire la distinction entre science et croyance – à propos de sujets scientifiques – au-delà des idées reçues [36] ;
- être en mesure de choisir et utiliser des solutions utilisant des algorithmes y compris d'intelligence artificielle, pour ses besoins et usages, de manière technocritique [27].

Les liens avec la recherche sont doubles :
- en sciences du numérique, permettre de renforcer les liens entre recherche et société [32,33], au service de l'égalité des chances [34], mais aussi développer des projets de recherche participative [35] rendus possibles par la popularisation des notions informatiques. On doit développer la capacité citoyenne de pouvoir poser des questions scientifiques dans ces domaines ;
- en sciences de l'éducation, former les enseignants aux nouveaux outils algorithmiques et utiliser ces outils pour étudier et modéliser les mécanismes d'apprentissage.

L'expérience démontre que c'est à travers les démarches participatives et créatives pour la programmation, définies dans l'article [22] et illustrées en Fig. 2, que l'adhésion et l'engagement de l'apprenant sont supérieurs à une simple formation transmissive. Dans ce cadre, il est essentiel de distinguer les activités d'apprentissage de la programmation des activités d'apprentissage par le biais de la programmation qui elles ont des applications qui dépassent l'étude des sciences du numérique (grammaire, narration interactive, écriture automatique, visualisation de données, etc.). Le propos de la pensée informatique étant celui-là même : outiller conceptuellement l'apprenant afin qu'il puisse déléguer, en parfaite conscience, des tâches à une machine et ainsi tirer le meilleur parti des technologies disponibles.

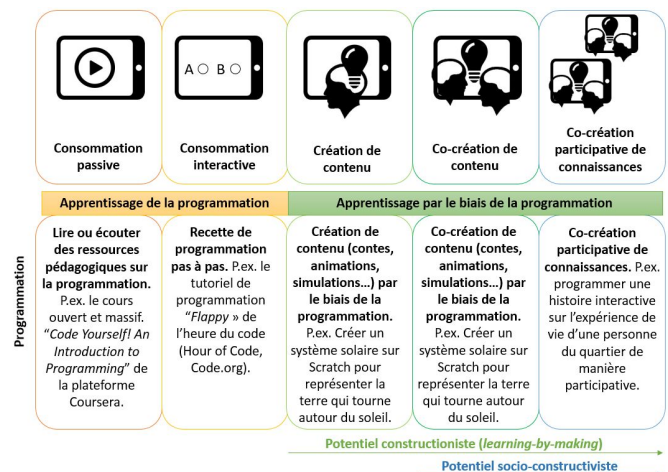

**Fig. 2.** Différences entre l'apprentissage de la programmation ou d'autres aspects de l'informatique de manière procédurale et de la programmation créative comme outil de modélisation de connaissances, repris de [22].

Concrètement, nous avons identifié 3 leviers :
1. **l'explicitation du terme intelligence artificielle** (*buzzword* surreprésenté dans les médias) qui motive les personnes à faire une démarche de découverte d'un pan de la « science des données » au sens défini dans l'article [24], hors disciplines STIM [23];
2. **la manipulation pratique qui** permet, même sur un temps court, d'établir la confiance nécessaire pour entrer dans le sujet, de transmettre une vision concrète des notions abordées et d'en soulever l'intérêt scientifique et culturel[10,13,20] ;

3. **l'apprentissage au fil de l'eau** qui permet de se former au quotidien, de manière progressive, modulaire et itérative. [15,20],

## 3   Une méthodologie pour mettre en œuvre ces objectifs

Une majorité de la population ignore ce qu'est la pensée informatique, est peu familière des notions de données, d'algorithmes, de programmes, de langages, et de réseaux, même si ce sont des « usagers de l'informatique » dans leurs domaines respectifs (scientifiques, industriels, médicaux, artistiques,...). Ils ont été formés à l'utiliser, rarement à comprendre les éléments fondateurs ainsi que l'étendue et les limites des possibilités permises par le numérique. Le besoin en formation, immense, est une situation quasi inédite depuis la mise en place de l'École obligatoire à la fin du XIX[e] siècle, comme le rappelle Gérard Berry [30]. De plus, le nombre de personnes en capacité de former les autres est largement insuffisant.

Cette situation particulière nécessite des réponses créatives dans les approches de formation, notamment : (i) en mixant les publics et les modalités, et en (ii) impliquant des personnes possédant des compétences variées et différents niveaux de connaissance, à la fois en tant que formateur·rice·s et publics (dans cette démarche, les deux rôles se mixent). Notre proposition sera d'inscrire cette démarche dans un continuum éducatif tout au long de la vie [31].

C'est en pensant les choses globalement et en agissant localement que des objectifs aussi ambitieux pourront être mis en place (voir [30] pour un descriptif très précis de propositions,de mesures concrètes et opérationnelles, répondant à des besoins effectifs et bien identifiés : http://tinyurl.com/yyen526u).

Concrètement, nous proposons de nous appuyer sur la méthodologie développée dans le cadre de Class'Code pour développer des ressources diversifiées. Elles seront le support d'une formation hybride, performative et diffusante au sens suivant : un parcours de formation en ligne librement accessible avec une certification gratuite et des moyens de partager et s'entraider, complété de propositions de temps de rencontre en ligne ou sur le territoire afin de prolonger ce partage et se mettre au service de projets locaux sur ces sujets ; La granularité et l'indexation de tous les contenus permettra de dériver et de construire facilement d'autres parcours.

Au niveau de la formation, le paradigme est le suivant : chaque apprenant (adulte, sénior, enfant) partage ce qu'il a appris avec son entourage – dans sa famille, son quartier, son association ou son entreprise : être capable de transmettre ce qui a été appris étant un objectif de formation en soi.

Le principe de ne pas segmenter les cibles, permet de partager les complémentarités de compétences : on a, par exemple, à la fois besoin des témoignages des usagers et des éléments d'expertise technique pour se forger une culture scientifique et technique commune citoyenne pertinente sur ces sujets. La quasi-totalité des publics sont à l'heure actuelle à un stade initial de formation. Même pour les personnes d'un niveau avancé ont les compétences sont parcellaires (on peut ainsi manquer de culture scientifique tout en sachant coder ou réfléchir au numérique sans compétence informatique).
Nous proposons une démarche partenariale et collective qui sera concrétisée à travers l'action collaborative de tous les partenaires. Tous les projets émergents dans les territoires ont vocation à être encouragés et soutenus. Les différentes expériences et pratiques feront l'objet d'une mutualisation et seront partagés dans toutes les strates sociales et sur l'ensemble du territoire.

C'est à travers la co-production de contenus que seront mis en oeuvre ce partage et cet apprentissage participatifs : par exemple en produisant des vidéos éducatives ou bien en dérivant des activités débranchées [16,17]. Les apprenants deviennent ainsi les acteurs de leur propre formation. Les ressources seront diffusées sous licence libre et ouverte et seront donc réutilisables et transformables. Cela permettra à toute personne physique ou morale d'utiliser ou de dériver des services marchands ou non-marchands basés sur ces ressources (formations spécifiques, animations de temps de loisir, etc.). Cela permettra aussi de mutualiser les ressources et les outils communs. Au delà, lier les initiatives associatives aux domaines économiques et à la formation professionnelle offre des leviers financiers nécessaires à la pérennisation de ces actions.
La création de cette structure qui permettra, de façon agile, de développer les produits et/ou services marchands est actuellement en cours d'élaboration.

A partir de cette vision globale, nous proposons une mise en œuvre locale : en se donnant pour objectif d'avancer ensemble collectivement sans aucun clivage d'âge, de secteur professionnel ou de segmentation entre public et privé ; en s'appuyant sur les acteurs de terrain et en travaillant avec eux à l'appropriation des outils, ressources et thématiques proposés, en co-construisant les ressources manquantes, en accompagnant leur mise en œuvre, en faisant progressivement monter en compétences un réseau de formateur·rice·s aux profils volontairement divers et polyvalents, en favorisant l'émergence de projets originaux, la remontée des besoins et des bonnes pratiques.

Nous proposons de nous mettre au service de ces personnes au niveau national sous forme de services d'ingénierie pédagogique, de formation et d'animation de réseaux :
- un partage de ressources ouvertes librement réutilisables pour toutes et tous,
- la recherche de ressources manquantes, la création de nouvelles et leur documentation,
- un accompagnement des formations présentielles de formateur·rice·s, avec la création de services formation encadrée,
- un partage de compétences en matière d'animation de communautés apprenantes, de réseaux hétérogènes de partenaires, y compris pour des besoins de montage de dossier, ou de promotion du débat citoyen sur ces questions

Beaucoup de ces éléments sont issus du projet Class'Code [20], ils ont donc été validés par un grand nombre et dans la durée. Ici, ils sont enrichis et élargis. Nous commençons d'ailleurs à développer avec les acteurs de la recherche, de l'éducation, des arts et en lien avec le monde économique et politique, des propositions de formation et de sensibilisation à même de toucher des publics variés sur les sujets que l'on rattache à l'intelligence artificielle. Ces expérimentations pilotes seront ensuite re-partagées.

Si le projet Class'Code est un point de départ de cette initiative, le projet de Petite École du Numérique [37] qui met la priorité sur le partage de ressources, la formation des formateur·rice·s et l'action territoriale, pourrait-il être un moyen d'aller vers cette ambition ?

## 4   Conclusion : vers une université citoyenne

Ce qui est proposé porte un nom : une *Université, Citoyenne en Sciences et Culture du Numérique (#UCscN)* qui s'inscrit dans la tradition des universités populaires[3]. Il s'agit donc très simplement d'étendre à toutes et tous cette éducation pour penser l'informatique en capitalisant sur l'expérience acquise de Class'Code en formant les professionnel·le·s de l'éducation. Une partie des contenus et des méthodes sont identiques et une partie du travail que nous faisons au niveau des partenaires que nous rassemblons pour s'en saisir et coopérer s'en trouve élargie, tandis que nous cherchons à décloisonner ces temps de formation en rassemblant au-delà des professionnel·le·s de l'éducation. C'est une démarche partenariale et collective implémentée à travers l'action collaborative de ses partenaires, en reprenant la méthode de Class'Code avec des objectifs élargis pour défendre une éducation citoyenne qui permet de maîtriser collectivement le numérique [11].

---
[3] On parle ici d'université au sens de Peirce, comme « une association d'hommes [et de femmes ] […] dotée et privilégiée par l'État, en sorte que le peuple puisse recevoir une formation [guidance] intellectuelle, donc au sens d'université populaire ».

**Remerciements.** Toute l'équipe et les partenaires impliqués : Educazur, Inria, Réseau Canopé, Kids Code Jeunesse, La Ligue de l'enseignement, LINE, Magic Makers, U. Côte d'Azur, U. Nantes et l'ensemble des partenaires de Class'Code.

**Annexe.** Les éléments opérationnels et organisationnels sont dans le document compagnon [31].

**Références** (les références seront ré-ordonnées dans la version finale du papier).